# Shock Loading Accesses Reaction Pathways beyond Accelerated Thermolysis

Mingjie Wen, Jiahe Han, Zihan Zhou, Hao Wang, Xueyong Guo, Qingzhao Chu*, Dongping Chen

*State Key Laboratory of Explosion Science and Safety Protection, Beijing Institute of Technology, Beijing 100081, P. R. China*

**Abstract:** Thermolysis and shock loading can produce similar major stable products in energetic materials, despite different chemical evolution and reaction pathways. Here, we develop EMFF-MACE, a reactive machine-learning potential for C/H/N/O energetic materials, enabling thermolysis simulations of 20 materials and shock simulations of 10 on a common potential energy surface. Comparisons at matched normalized reaction progress reveal divergent chemical states and local structural environments. Mean reaction-state similarity decreases from 94.0% at 20% progress to 25.9% at trajectory endpoints, while 47.1-92.6% of shock local environments lie outside corresponding material-specific thermolysis reference domains. Structural and reaction-network analyses reveal enhanced intermolecular reorganization and additional association and rearrangement branches under shock loading, despite partially shared early decomposition chemistry. These findings show that loading history influences the structural states and subsequent reaction pathways accessed during decomposition, rather than merely the rate of thermolysis.

* Corresponding author, E-mail: chuqz@bit.edu.cn (Q. Chu).

## Introduction

Energetic materials can release substantial chemical energy through thermal decomposition or shock-induced reactions (*1, 2*). Thermolysis is driven primarily by sustained thermal excitation, whereas shock loading involves rapid compression, structural deformation, and heating. Experimental investigations and reactive molecular dynamics simulations have established the sensitivity of reaction initiation and chemical evolution to these conditions (*3, 4*). Although these loading modes exhibit distinct reaction behaviors, their differences cannot be characterized by reaction timing alone. A central issue is whether loading history also influences the chemical states and subsequent reaction pathways accessed during decomposition.

A major difficulty in such comparisons is that the two loading modes evolve under different thermodynamic conditions and timescales. Comparisons at identical simulation times may correspond to different stages of decomposition, while similar stable products, such as $N_2$, $CO_2$, and $H_2O$, can mask differences in intermediate chemistry. A meaningful comparison therefore requires a common reactive potential energy surface, a consistent measure of reaction progress, and comparable descriptions of chemical species and local atomic environments. Advances in machine-learning potentials (MLPs) have substantially extended the spatial and temporal scales accessible to reactive molecular dynamics simulations. Deep-potential models enabled simulations of shock-induced reactions and Chapman-Jouguet (CJ) detonation states (*5, 6*), while transferable C/H/N/O potentials extended reactive simulations to multiple energetic compounds (*7*). Our previous EMFF-2025 model further established a general description of crystal structures and thermal decomposition across 20 C/H/N/O energetic materials (*8*). Full chemical names and abbreviations of all 20 energetic materials are listed in Table S1. Fig. 1(a) summarizes the annual number of selected energetic-material MLP studies from 2020 to 2026, with their corresponding material systems documented in Table S2. Fig. 1(b) compares the reported maximum temperatures and pressures of representative general-purpose and energetic-material potentials, highlighting the extension of thermodynamic sampling and the development of the EMFF series. These advances have expanded the range of materials and thermodynamic conditions accessible to reactive simulations. However, broader model coverage alone does not establish whether different loading histories change only the timing of reactions or also their subsequent chemical evolution.

Differences in reaction initiation do not necessarily capture how reactive intermediates subsequently transform under different loading conditions. Reactive molecular dynamics studies of 1,3,5-trinitroperhydro-1,3,5-triazine (RDX) showed that temperature and density affect decomposition and product formation, while shock loading can initiate rapid chemical transformations (*9, 10*). Studies of pentaerythritol tetranitrate (PETN) revealed orientation-dependent shock initiation and collision-induced activation (*11, 12*), whereas investigations of 2,4,6-Trinitrobenzene-1,3,5-triamine (TATB) identified the importance of intramolecular strain and local structural environments in mechanochemical reactions (*13, 14*). Further studies of other molecular and ionic energetic materials demonstrated the effects of density, precompression, and structural anisotropy on their chemical responses (*15-19*). Collectively, these studies clarify how loading conditions influence initial chemical activation, while the subsequent evolution of reactive intermediates remains less well characterized. For thermal decomposition, EM-HyChem (*20*), developed from the hybrid chemistry (HyChem) framework for hydrocarbon fuels (*21-24*), describes energetic-material pyrolysis and subsequent oxidation through fragment-based kinetics. Under shock loading, experimental and theoretical studies of RDX revealed accelerated decomposition and proposed radical-mediated cyclic reactions (*25, 26*). Shear-induced self-accelerating chemistry has also been proposed for energetic crystals (*27*). These studies establish the importance of loading conditions in reaction initiation and intermediate chemistry. However, the extent to which loading-dependent intermolecular reorganization accompanies changes in subsequent reaction pathways remains less directly established, particularly when different loading modes share early decomposition chemistry and major stable products.

Here, we systematically compare thermolysis and shock-induced chemistry to determine whether loading-dependent differences extend beyond reaction timing to subsequent chemical evolution. We develop EMFF-MACE, a reactive machine-learning interatomic potential based on the message-passing atomic cluster expansion (MACE) architecture (*28*) and the EMFF-2025 dataset (*8*), enabling thermolysis simulations of 20 C/H/N/O energetic materials and shock simulations of 10 representative systems on a common potential energy surface. Comparisons at matched normalized reaction progress reveal increasingly distinct chemical states and local structural environments, despite the formation of similar major stable products. Analyses of intermolecular proximity, connectivity, and representative reaction networks further identify additional association and rearrangement branches under shock loading. Together, these findings show that loading history influences accessible structural states and subsequent reaction pathways, rather than merely the timing of thermolysis.

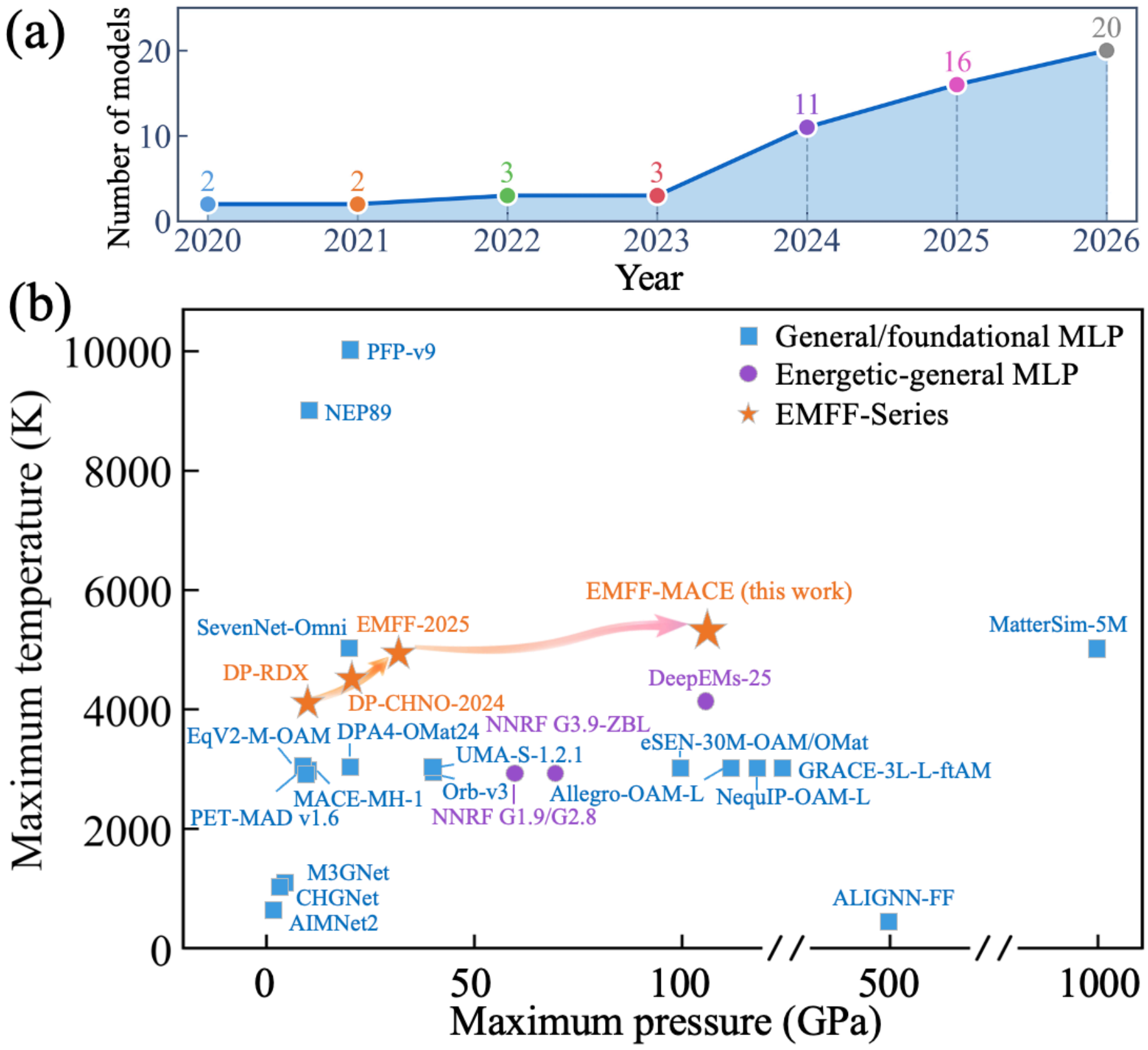

**Fig. 1.** Development of machine-learning potentials (MLPs) for energetic materials and their reported temperature-pressure coverage. (a) Annual number of selected energetic-material MLP studies from 2020 to 2026, including this work (Table S2). (b) Maximum reported temperatures and pressures of representative general-purpose MLPs (blue squares), energetic-material MLPs (purple circles), and the EMFF series and its predecessors (orange stars). Orange arrows trace the development toward EMFF-MACE. Literature sources are provided in Tables S2 and S3. Reported limits indicate explored or documented conditions rather than uniformly validated accuracy ranges

# Results

### Accuracy and applicability of EMFF-MACE

EMFF-MACE was first evaluated against density functional theory (DFT) reference energies and atomic forces on the test set. In Fig. 2(c), the predictions closely follow the diagonal. The energy mean absolute error (MAE) and root-mean-square error (RMSE) are 18.68 and 23.78 meV·atom$^{-1}$, respectively, with a coefficient of determination ($R^2$) of 0.9964. The corresponding force MAE and RMSE are 0.368 and 0.527 eV·Å$^{-1}$, respectively, with $R^2$ of 0.9740. The material-resolved energy and force errors vary across the 20 systems (Fig. S1). Because thermolysis and shock involve substantial temperature increases and volumetric deformation, the model was further tested for RDX, HMX, and CL-20 over 300-4000 K and linear cell-scaling factors of 0.80-1.20 (Fig. S2). Energy and force errors remain low near equilibrium and increase smoothly with deformation, with the largest deviations under strong compression. At a scaling factor of 0.80, the maximum energy and force MAEs are 58 meV·atom$^{-1}$ and 2.07 eV·Å$^{-1}$, respectively.

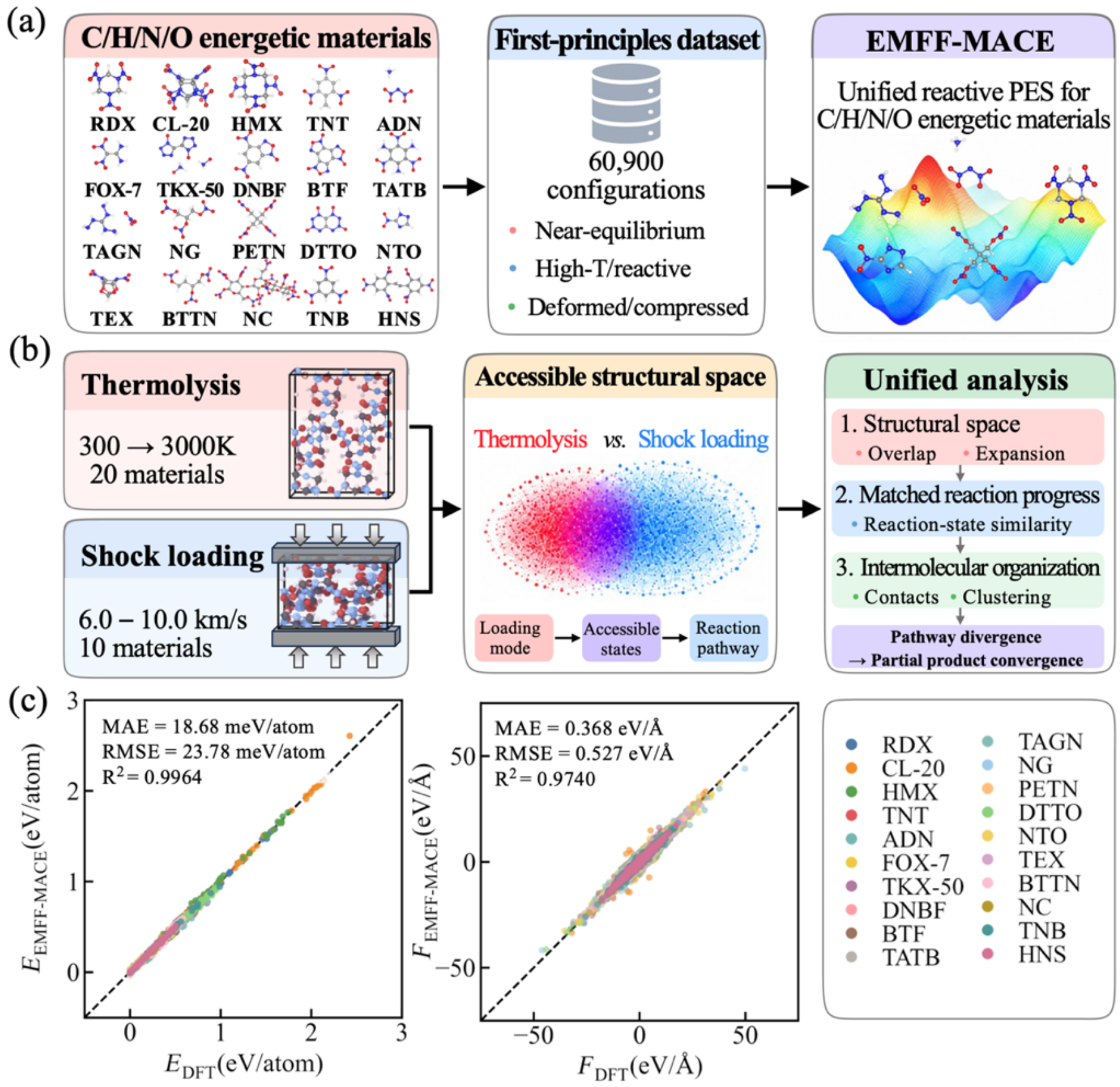

**Fig. 2.** Construction and validation of EMFF-MACE and the unified framework for comparing thermolysis and shock loading. (a) First-principles training dataset for 20 C/H/N/O energetic materials, including additional strongly deformed configurations of RDX, HMX, and CL-20 (Table S4). (b) Programmed-heating thermolysis and multiscale shock technique (MSST) simulations on the same potential energy surface, followed by matched-progress comparison of chemical states, accessible structural space, and pathway evolution. (c) EMFF-MACE energies and atomic forces compared with DFT references.

Crystal structures and equations of state were then used to assess the condensed-phase response of the model (Fig. S3). The mean absolute errors are 1.22% for lattice parameters and 2.16% for unit-cell volume. EMFF-MACE also reproduces the DFT energy-volume trends of RDX, HMX, PETN, and TATB. Across all 20 materials, the mean energy deviation is 0.015 eV·atom$^{-1}$ over the volume-ratio range $V/V_0$ = 0.779-1.260, increasing to 0.57 eV·atom$^{-1}$ at $V/V_0$ = 0.512 under extreme compression.

Representative bond-length, bond-angle, and dihedral scans in RDX and PETN further show that EMFF-MACE reproduces the principal DFT minima, curvatures, and periodic barriers, with the largest deviations confined to strongly stretched configurations (Fig. S4). Together, these tests support the applicability of EMFF-MACE to the evaluated equilibrium, high-temperature, compressed, and locally distorted configurations, while identifying the largest deviations under extreme compression and strong bond stretching.

**Thermal decomposition and product evolution**

We investigated the thermolysis of 20 C/H/N/O energetic materials using EMFF-MACE under a consistent programmed-heating protocol (Fig. 3). In Fig. 3(a), the normalized reactant fractions exhibit pronounced material-dependent differences in reaction-onset temperature and the temperature interval of reactant consumption. NC and BTTN undergo substantial reactant depletion at relatively low temperatures, whereas FOX-7 and TATB retain high reactant fractions until much higher temperatures. Previous comparative thermogravimetry and differential scanning calorimetry (TG/DSC) measurements of PETN, RDX, HMX, and HNS also revealed pronounced material-dependent decomposition behavior (*29*), qualitatively supporting the differences observed in the present simulations.

To examine the associated chemical evolution, RDX, PETN, FOX-7, and TKX-50 were selected as representative systems with different molecular structures and decomposition characteristics (Fig. 3(b)). During RDX decomposition, nitrogen-oxygen intermediates, including $NO_2$, NO, and $N_2O$, appear at early stages, followed by the accumulation of $N_2$, $H_2O$, and $CO_2$, reaching final normalized abundances of 45.6%, 34.6%, and 18.5%, respectively. The transient formation and subsequent conversion of nitrogen-containing intermediates are broadly consistent with previous in situ Fourier-transform infrared spectroscopy observations of RDX thermolysis (*30*). For PETN, $H_2O$ and $CO_2$ account for 38.9% and 37.8% of the final normalized species abundance, respectively, followed by $N_2$ at 19.1%, while $NO_2$ and NO appear mainly as transient intermediates. Previous thermal analysis and mass spectrometry studies have also identified $NO_2$ formation as an important feature of PETN decomposition (*31*). For FOX-7, the formation of $N_2$, $H_2O$, and $CO_2$ increases progressively during reactant consumption, reaching final normalized abundances of 50.7%, 31.2%, and 17.9%, respectively. The evolution of nitrogen- and oxygen-containing intermediates is consistent with previously reported decomposition characteristics (*32*). In contrast, the nitrogen-rich ionic material TKX-50 exhibits early formation of $NH_2OH$ and nitrogen-containing intermediates, followed by the accumulation of $N_2$ and $H_2O$, with final normalized abundances of 68.0% and 32.0%, respectively. These intermediates are consistent with previously reported thermal decomposition chemistry involving proton transfer and N–N/C–N bond transformations (*33*). The complete species evolution of the remaining materials is provided in Fig. S5.

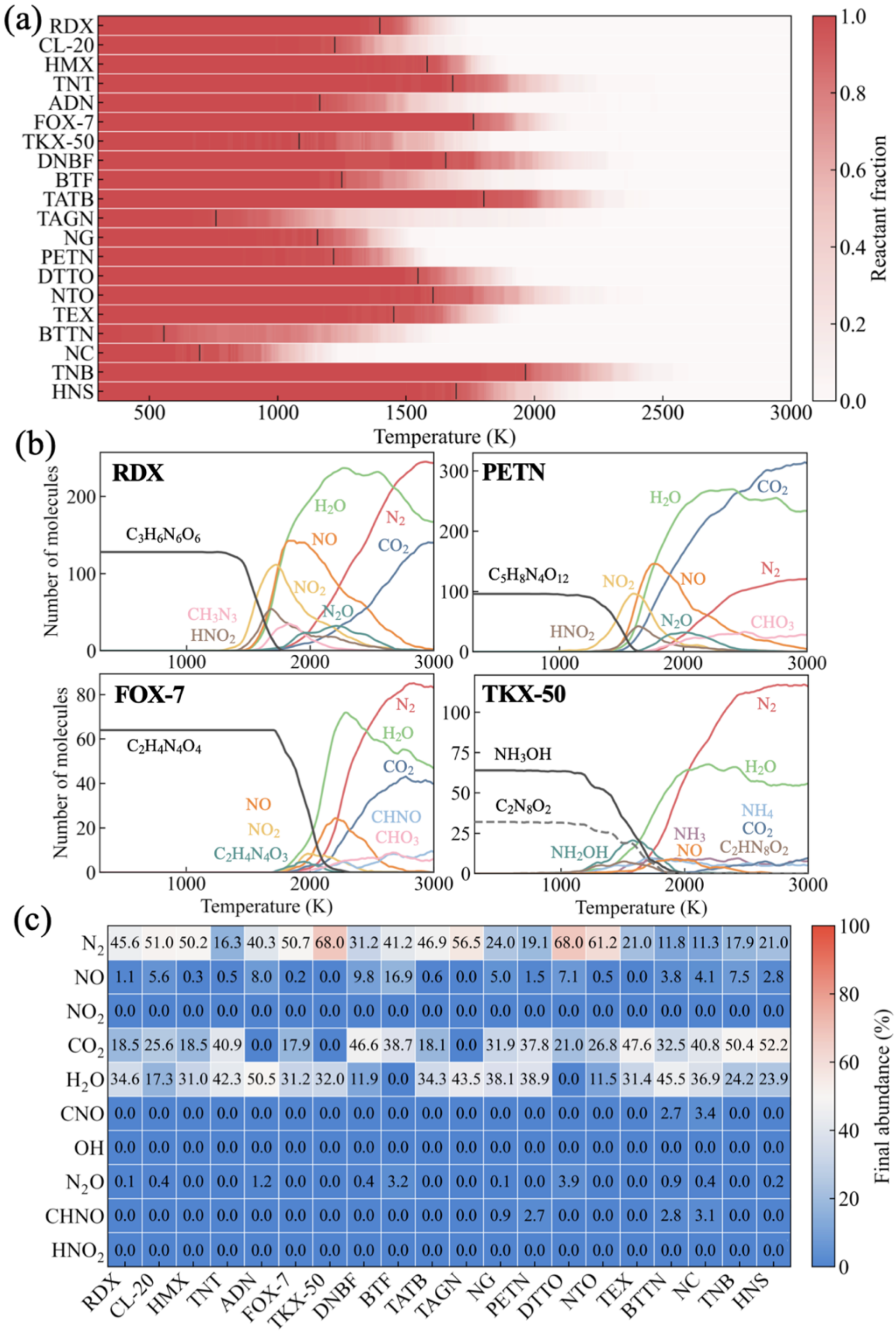

**Fig. 3.** Reactant consumption and product evolution during programmed thermolysis of 20 energetic materials. (a) Normalized reactant fraction as a function of temperature. Black ticks indicate reaction-onset temperatures, defined by a 5% decrease from the initial reactant fraction. (b) Temperature-dependent evolution of representative reactants, intermediates, and small-molecule

products in RDX, PETN, FOX-7, and TKX-50. (c) Normalized abundances of the major final product species across the 20 materials.

Despite these differences in reactant consumption and intermediate evolution, the 20 energetic materials exhibit a common tendency toward the formation of major small-molecule products (Fig. 3(c) and Fig. S6). $N_2$, $CO_2$, and $H_2O$ dominate the final product distributions, with their combined normalized abundance averaging 94.9% across the 20 materials and reaching at least 79.9% in every system. $N_2$ is particularly abundant in TKX-50 and DTTO (68.0% each), $CO_2$ dominates in HNS and TNB (52.2% and 50.4%), and $H_2O$ reaches 50.5% in ADN and 45.5% in BTTN. In contrast, $NO_2$ is nearly absent from the final states, while NO persists only in selected systems, indicating the predominantly intermediate character of these nitrogen-oxygen species.

The predominance of common small-molecule products is consistent with previous observations of shared high-temperature chemical environments in EMFF-2025 (*8*) and the fragment-based description of decomposition chemistry in EM-HyChem (*20*). However, similarity in final product composition does not establish equivalence in the intermediate chemical states or reaction pathways through which these products are formed.

**Shock response and detonation states**

To investigate the structural and thermodynamic responses of energetic materials under dynamic compression, we performed MSST simulations of 10 representative systems using the same EMFF-MACE potential energy surface. RDX was first examined over shock velocities of 6.0-10.0 $km \cdot s^{-1}$ by monitoring the evolution of relative volume, temperature, and pressure (Figs. 4(a)-(c)). At 6.0-7.0 $km \cdot s^{-1}$, RDX undergoes rapid compression and remains in a relatively stable high-pressure state, with only limited subsequent changes in $V/V_0$, temperature, and pressure. At 7.6-8.0 $km \cdot s^{-1}$, RDX exhibits an induction period following initial compression, followed by a sustained temperature rise, gradual volume recovery, and pressure relaxation toward the post-reaction state. The transition shifts to earlier times with increasing shock velocity: a clear induction period remains at 7.8 $km \cdot s^{-1}$, whereas major chemical transformation occurs earlier at 7.9 $km \cdot s^{-1}$. At 9.0-10.0 $km \cdot s^{-1}$, the system rapidly enters a high-temperature, high-pressure reactive state after compression.

The CJ state was determined from the tangency between the Rayleigh line and the reacted Hugoniot, using dynamically stabilized post-reaction states without sustained drift in volume, temperature, or pressure (*5, 34, 35*). As shown in Fig. 4(d), the trajectories at 7.8 and 7.9 km·s$^{-1}$ exhibit distinct reaction responses near the reaction-transition regime. The resulting RDX CJ state is characterized by a detonation velocity ($D_{CJ}$) of 7.92 km·s$^{-1}$, pressure ($P_{CJ}$) of 30.28 GPa, temperature ($T_{CJ}$) of 2285 K, and density ($\rho_{CJ}$) of 2.45 g·cm$^{-3}$. For comparison, the reactive force field (ReaxFF)-based Reactive Dynamics-to-CJ (Rx2CJ) approach predicts $D_{CJ}$ = 8.27 km·s$^{-1}$ and $P_{CJ}$ = 28.62 GPa (*36*), while the experimental reference values are 8.60 km·s$^{-1}$ and 33.79 GPa (*37*). The predicted detonation velocity is lower than both reference values, whereas the CJ pressure remains comparable to the reported results.

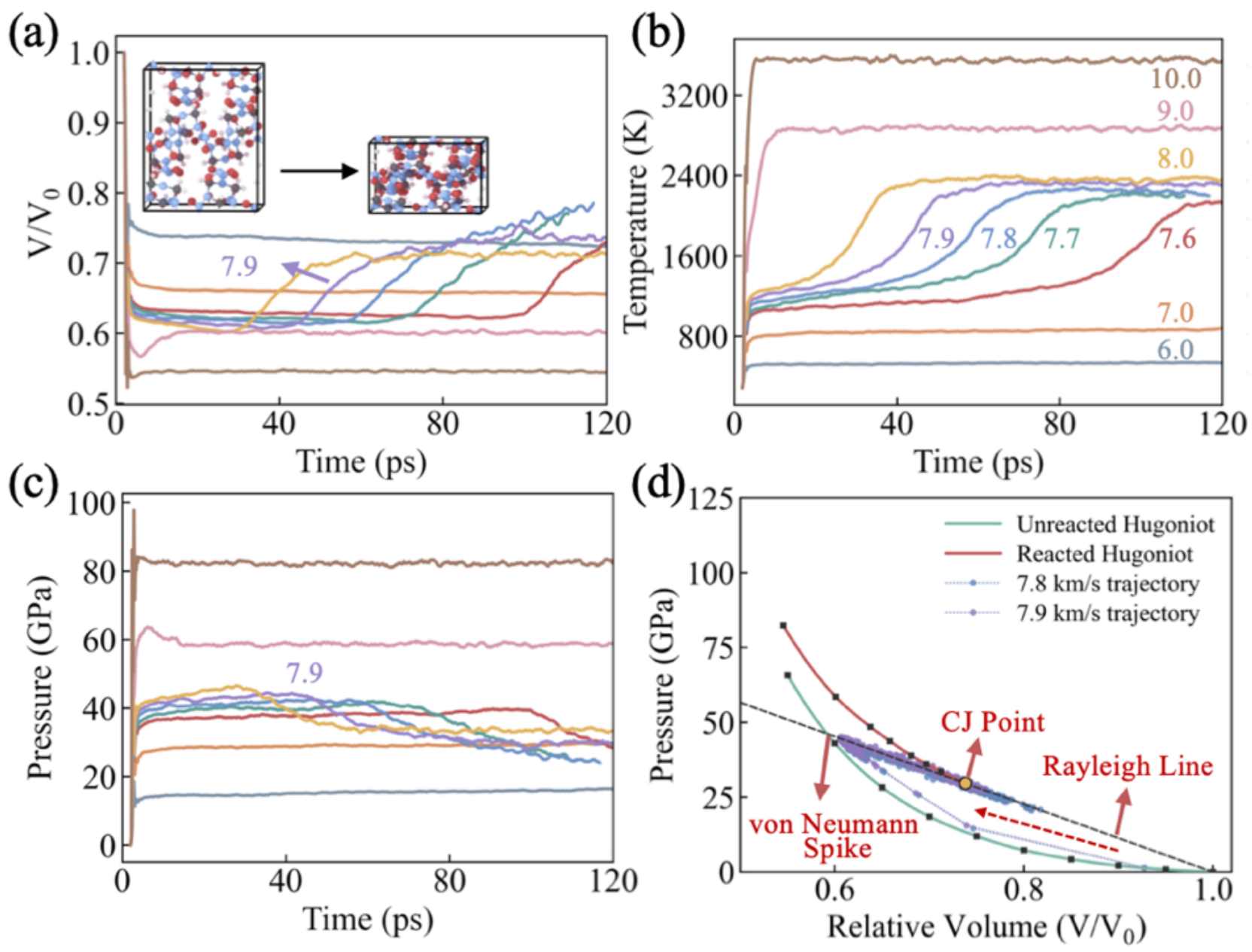

**Fig. 4.** Shock response of RDX and determination of the Chapman-Jouguet state. Time evolution of (a) relative volume $V/V_0$, (b) temperature, and (c) pressure at different shock velocities; the inset in (a) illustrates the structures before and after shock compression. (d) Unreacted and reacted Hugoniots, representative trajectories at 7.8 and 7.9 km·s$^{-1}$, and the CJ Rayleigh line. The CJ state is identified by the tangency between the Rayleigh line and the reacted Hugoniot, while its intersection with the unreacted Hugoniot defines the von Neumann state.

The same MSST and CJ-state analysis was then applied to HMX, PETN, FOX-7, NTO, BTF, CL-20, HNS, TNT, and TKX-50 (Table 1). The predicted $D_{CJ}$ values span 6.80-9.61 km·s$^{-1}$ across the 10 materials, from HNS and TNT at the lower end to CL-20 and TKX-50 at the higher end. Among the CJ parameters, DCJ shows the most consistent agreement with selected literature reference values, with a mean absolute relative deviation of 6.1% for the corresponding comparisons in Table 1. For example, the predicted $D_{CJ}$ values of PETN (7.32 km·s$^{-1}$), NTO (7.80 km·s$^{-1}$), and CL-20 (8.81 km·s$^{-1}$) are close to reported Rx2CJ or machine-learning-potential results of 7.44, 7.90, and 9.20 km·s$^{-1}$, respectively (*5, 6, 36*). The predicted detonation velocity of TKX-50 (9.61 km·s$^{-1}$) is also close to the reported EXPLO5 values of 9.64-9.70 km·s$^{-1}$ (*38, 39*).

Compared with $D_{CJ}$, the post-reaction pressure, temperature, and density are more sensitive to initial density, reaction extent, and final-state description. The predicted $P_{CJ}$ values for RDX and PETN, 30.28 and 24.27 GPa, are close to the corresponding Rx2CJ/ReaxFF values of 28.62 and 22.47 GPa (*36*). For BTF, the predicted $P_{CJ}$ of 32.86 GPa agrees closely with the experimental range of 33.8-34.5 GPa (*40*). For CL-20, the predicted value of 44.15 GPa closely agrees with the reported RMD-QMMD result of 44.24 ± 1.15 GPa (*19*), further supporting the model's ability to reproduce representative high-pressure detonation states. For systems with independent CJ-density references, the predicted $\rho_{CJ}$ values for RDX, PETN, and CL-20 are 2.45, 2.31, and 2.81 g·cm$^{-3}$, respectively, consistent in magnitude with previous Rx2CJ, Becker-Kistiakowsky-Wilson (BKW) equation-of-state, and machine-learning-potential results (*5, 36, 41*). Relative to the selected external reference values, the mean absolute relative deviations are 26.9% for $P_{CJ}$ and 9.6% for $\rho_{CJ}$. Larger pressure deviations are observed for FOX-7, NTO, and TKX-50. Overall, the predicted CJ pressures capture substantial material-dependent variation, although their quantitative agreement with reference data is less consistent than that of the detonation velocities.

The comparison of CJ temperatures reveals differences between the MSST simulations and equilibrium thermochemical calculations. The predicted values for FOX-7 and CL-20, 1938 and 2436 K, are close to reported reactive molecular dynamics and Deep Potential results of 1960 and 2500 K, respectively (*5, 42*). As an independent equilibrium thermochemical reference, EXPLO5 calculations were also performed for all 10 materials (Fig. S7). For RDX, EXPLO5 gives $D_{CJ}$ = 8.82 km·s$^{-1}$, $P_{CJ}$ = 34.14 GPa, and $T_{CJ}$ = 3738 K, consistent with previously reported EXPLO5 values of 8.80 km·s$^{-1}$, 34.9 GPa, and 3739 K (*43*). For CL-20, the predicted $D_{CJ}$ and $P_{CJ}$ are 9.75 km·s$^{-1}$ and 44.83 GPa, respectively, also within the reported EXPLO5 ranges of 9.63-9.77 km·s$^{-1}$ and 44.21-44.85 GPa (*44*). EXPLO5 determines detonation states under chemical-equilibrium assumptions, whereas the MSST simulations describe the coupled evolution of compression, chemical reactions, and energy redistribution over finite simulation timescales. Consequently, the MSST-derived temperatures correspond to dynamically accessed post-reaction states rather than equilibrium thermochemical states. Differences in reaction extent and product distributions therefore contribute to the distinct descriptions provided by the two approaches. These differences highlight the importance of accounting for reaction progress and product composition when comparing atomistic simulations with equilibrium thermochemical calculations.

**Table 1.** Chapman-Jouguet parameters of ten energetic materials predicted by EMFF-MACE and compared with EXPLO5 and literature data.

| Material[a] | $\rho_0$ (g·cm$^{-3}$) | $D_{CJ}$ (km·s$^{-1}$) | $P_{CJ}$ (GPa) | $T_{CJ}$ (K) | $\rho_{CJ}$ (g·cm$^{-3}$) | Source [b] |
|---|---|---|---|---|---|---|
| RDX | 1.81 | 7.92 | 30.28 | 2285 | 2.45 | EMFF-MACE (this work) |
| | | 8.82 | 34.14 | 3738 | 2.39 | EXPLO5 (this work) |
| | 1.76 | 8.27 | 28.62 | 2700 | 2.31 | Rx2CJ (*36*) |
| | 1.77 | 8.60 | 33.79 | — | — | Experiment (*37*) |
| | 1.81 | 8.80 | 34.90 | 3739 | — | EXPLO5 (*43*) |
| HMX | 1.90 | 8.61 | 44.21 | 2266 | 2.83 | EMFF-MACE (this work) |
| | | 9.18 | 37.89 | 3667 | 2.49 | EXPLO5 (this work) |
| | 1.86 | 8.40 | 32.68 | 2680 | 2.48 | Rx2CJ (*36*) |
| | 1.89 | 9.11 | 39.5 | — | — | Experiment (*37*) |
| | 1.90 | 9.23(calc.) /9.10(exp.) | 38.0 | — | — | Calculation / experiment (*45*) |
| PETN | 1.69 | 7.32 | 24.27 | 2758 | 2.31 | EMFF-MACE (this work) |
| | | 8.11 | 27.87 | 4058 | 2.25 | EXPLO5 (this work) |
| | 1.72 | 7.44 | 22.47 | 2460 | 2.25 | Rx2CJ (*36*) |
| | 1.67 | 7.98 | 30.60 | — | — | Experiment (*37*) |
| | 1.77 | 8.42 | 31.80 | 2833 | 2.43 | BKW-EOS (*41*) |
| | 1.78 | 8.47 | 31.30 | 4038 | | EXPLO5 (*46*) |
| FOX-7 | 1.90 | 8.61 | 46.92 | 1938 | 2.97 | EMFF-MACE (this work) |
| | | 8.83 | 33.94 | 3107 | 2.46 | EXPLO5 (this work) |
| | 1.85 | 8.42 | 35.40 | 1960 | 2.53 | RxMD-QMMD (*42*) |
| | 1.88 | 8.87 | 34.00 | — | — | Literature benchmark (*47*) |
| NTO | 1.88 | 7.80 | 42.88 | 1715 | 3.01 | EMFF-MACE (this work) |
| | | 8.27 | 28.29 | 2898 | 2.41 | EXPLO5 (this work) |
| | 1.90 | 7.90 | 33.00 | — | — | Deep Potential + MSST (*6*) |
| | 1.93 | 8.70 | 34.90 | — | — | Literature benchmark (*48*) |
| | 1.92 | 8.45 | 32.90 | — | — | K-J/EXPLO5 (*49*) |
| BTF | 1.90 | 7.40 | 32.86 | 3215 | 2.78 | EMFF-MACE (this work) |
| | | 8.96 | 36.26 | 4836 | 2.49 | EXPLO5 (this work) |
| | 1.82 | 8.39 | 33.80 | 3990 | — | Experiment (*40*) |
| | 1.84 | 8.44 | 34.50 | 4170 | — | Experiment (*40*) |
| | 1.85-1.86 | 8.49 | 33.00-36.00 | 3700-5000 | — | Literature range (*40*) |
| CL-20 | 2.03 | 8.81 | 44.15 | 2436 | 2.81 | EMFF-MACE (this work) |
| | | 9.75 | 44.83 | 4096 | 2.65 | EXPLO5 (this work) |
| | 2.03 | 9.20 | 50.30 | 2500 | 2.76 | Deep Potential + MSST (*5*) |
| | 2.04 | 9.46 | 46.70 | — | — | EXPLO5 (*50*) |
| | 2.00 | 9.10 ± 0.23 | 44.24 ± 1.15 | 2687 ± 293 | | RMD-QMMD (*19*) |
| | 2.04 | 9.63-9.77 | 44.21 - 44.85 | — | — | K-J/EXPLO5 (*44*) |
| HNS | 1.74 | 6.80 | 30.04 | 2678 | 2.91 | EMFF-MACE (this work) |
| | | 7.14 | 21.89 | 3510 | 2.30 | EXPLO5 (this work) |
| | 1.74 | 7.50 | 23.64 | 3722 | — | Literature calculation (*51*) |
| | 1.74 | 7.61 | 24.30 | — | — | EXPLO5 (*52*) |
| | 1.74 | 7.41 | 24.10 | 3059 | — | BKW-EOS (*53*) |
| TNT | 1.70 | 7.01 | 32.42 | 2507 | 2.77 | EMFF-MACE (this work) |
| | | 6.96 | 20.36 | 3201 | 2.26 | EXPLO5 (this work) |
| | 1.70 | 7.16 | 22.02 | — | — | K-J (*54*) |
| | 1.63 | 6.93 | 21.00 | — | — | Experiment (*37*) |
| TKX-50 | 1.92 | 9.61 | 67.64 | 1561 | 3.17 | EMFF-MACE (this work) |
| | | 10.16 | 42.70 | 3479 | 2.45 | EXPLO5 (this work) |
| | 1.89 | 9.64 | 37.00 | — | — | EXPLO5 (*39*) |
| | 1.92 | 9.70 | 42.40 | — | — | EXPLO5 (*38*) |

[a] The crystal forms and initial densities of the ten materials used in the shock simulations are listed in Table S5.

[b] Reference values from experiments, reactive molecular dynamics, machine-learning potentials, BKW/K-J calculations, and EXPLO5 are included for comparison of CJ parameter magnitudes and cross-material trends. K-J denotes the Kamlet-Jacobs method.

EMFF-MACE captures the material-dependent shock responses and variations in CJ detonation velocity across the 10 energetic materials. The reactant-consumption trajectories also reveal distinct reaction timescales under the simulated shock conditions (Fig. S8). These results motivate a direct comparison of thermolysis and shock chemistry beyond their different reaction timescales.

**Loading-dependent divergence in chemical evolution**

To determine whether shock-induced chemistry differs from thermolysis beyond reaction timing, we first examined the species-level evolution of RDX, PETN, FOX-7, and TKX-50 under shock loading (Fig. 5(a)), with results for all 10 materials provided in Fig. S9. In RDX, $N_2$ becomes a major product following initial reactant depletion, accompanied by the formation of $CO_2$ and $H_2O$ and the transient appearance of nitrogen-oxygen intermediates. PETN exhibits relatively early $CO_2$ formation, followed by continued accumulation of $H_2O$ and $N_2$, while NO, $NO_2$, and $HNO_2$ appear predominantly as transient intermediates. FOX-7 shows pronounced late-stage increases in $N_2$ and $H_2O$, whereas TKX-50 exhibits substantial nitrogen-containing intermediate chemistry alongside sustained $N_2$ and $H_2O$ formation.

The product distributions at the trajectory endpoints exhibit common compositional features across the 10 energetic materials, despite differences in their preceding chemical evolution (Fig. 5(b)). Identified molecules and fragments were classified into $N_2$, $CO_x$, $H_2O$/OH, $NO_x$, $CH_xO_y$, $CH_xNO_y$, and other product families, with their abundances normalized within each material. $N_2$, $CO_x$, and $H_2O$/OH dominate the final product distributions in most systems, although their relative contributions vary substantially among materials. $N_2$ accounts for 67.3% in NTO and 49.5% in CL-20, while $CO_x$ reaches 50.7% in BTF and 36.4% in HNS. $H_2O$/OH contributes 39.1% in TKX-50. Appreciable fractions of $CH_xO_y$ and $CH_xNO_y$ species remain in PETN, HMX, TNT, and HNS at the final simulated states, indicating the persistence of chemically diverse molecular fragments. The corresponding species-resolved final product distributions are provided in Fig. S10.

Comparison with EXPLO5 shows that $N_2$, $CO_x$, and $H_2O$/OH remain prominent equilibrium product families for most materials, although their relative abundances differ substantially from the EMFF-MACE results (Figs. 5(b) and 5(c)). EMFF-MACE describes the species populations present at the finite-time trajectory endpoints, whereas EXPLO5 provides product distributions under chemical-equilibrium assumptions. The resulting differences in product-family abundances reflect the distinct states represented by the two approaches. The prevalence of $N_2$, $CO_2$, and $H_2O$ is also consistent with previous shock-reaction studies of RDX, CL-20, NTO, and FOX-7 (*5, 6, 10, 55*).

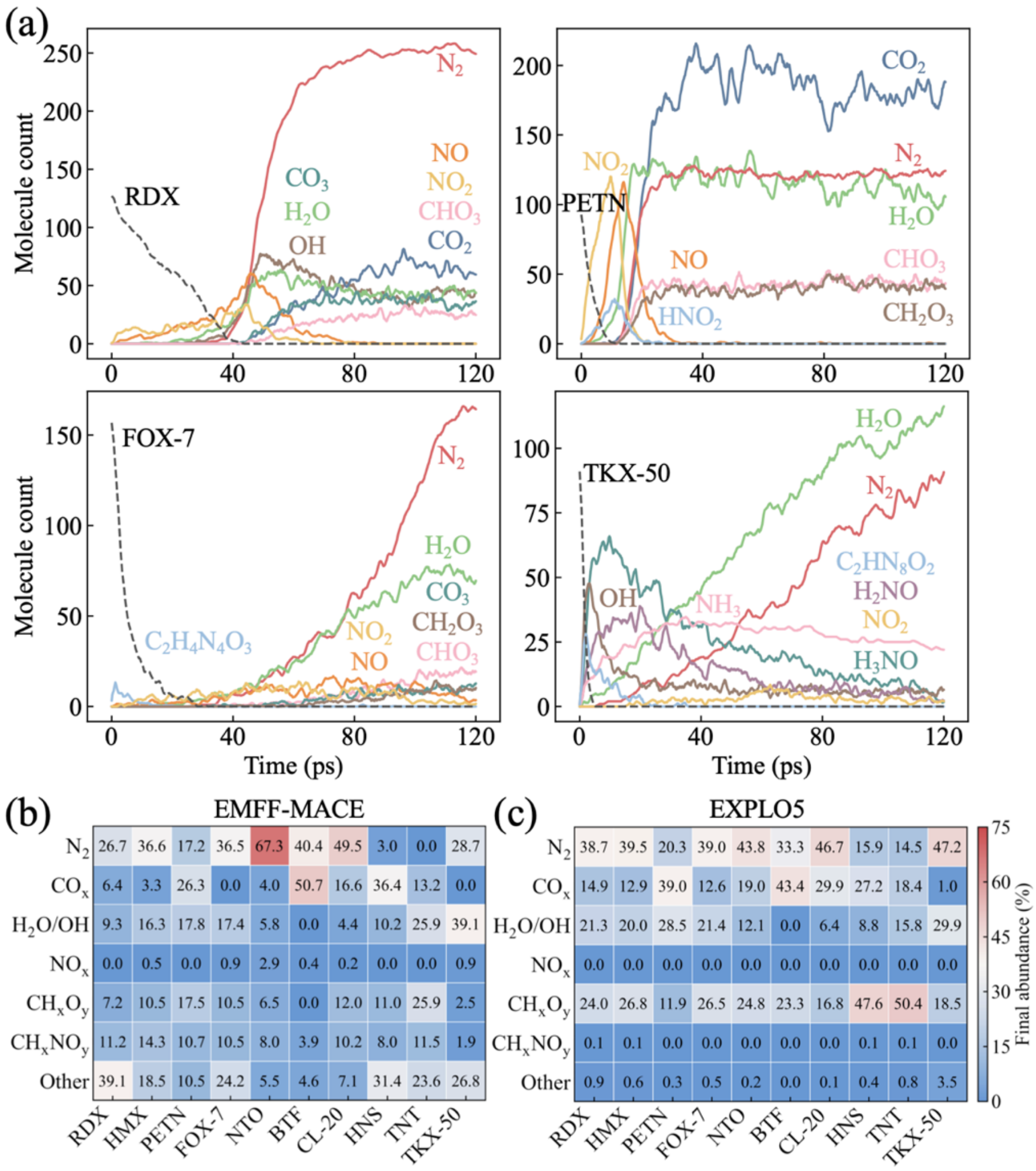

**Fig. 5.** Species evolution and final product distributions under shock loading. (a) Time evolution of representative reactants, intermediates, and products in RDX, PETN, FOX-7, and TKX-50 under representative shock conditions. Dashed lines denote the initial reactants or reactant components. (b) Normalized abundances of final product families predicted by EMFF-MACE for 10 energetic materials. (c) Corresponding equilibrium product-family distributions calculated using EXPLO5. The product families include $N_2$, $CO_x$, $H_2O/OH$, $NO_x$, $CH_xO_y$, $CH_xNO_y$, and other species.

Although thermolysis and shock loading produce similar major stable products, their intermediate chemical evolution differs substantially. To compare the two loading modes independently of their different reaction timescales, we aligned the trajectories by normalized reaction progress, defined relative to the final parent-reactant conversion reached in each trajectory (Section S1.1). Reaction-state similarity was then quantified from the overlap between their normalized species-abundance distributions at matched progress (Section S1.2 and Fig. 6(c)).

The temperature and pressure histories of thermolysis and shock loading exhibit distinct characteristics when expressed as functions of normalized reaction progress (Figs. 6(a) and 6(b)). Across the 10 energetic materials, thermolysis generally reaches higher temperatures at equivalent normalized progress, whereas shock loading produces substantially higher pressures. This difference reflects the distinct thermodynamic histories of the two loading modes: thermolysis proceeds under sustained heating, whereas substantial reactant consumption under shock loading is observed at comparatively lower temperatures and substantially higher pressures. When compared at corresponding normalized reaction progress, thermolysis and shock loading also exhibit increasing differences in their species-resolved chemical states (Fig. 6(c)). Averaged over the 10 energetic materials, the reaction-state similarity decreases from 94.0% at 20% reaction progress to 86.6%, 73.6%, 53.2%, and 25.9% at 40%, 60%, 80%, and 100% reaction progress, respectively. The two loading modes exhibit substantial chemical-state overlap at early stages, but their species compositions increasingly diverge at matched normalized reaction progress, demonstrating that reaction progress alone does not uniquely characterize chemical evolution across loading histories.

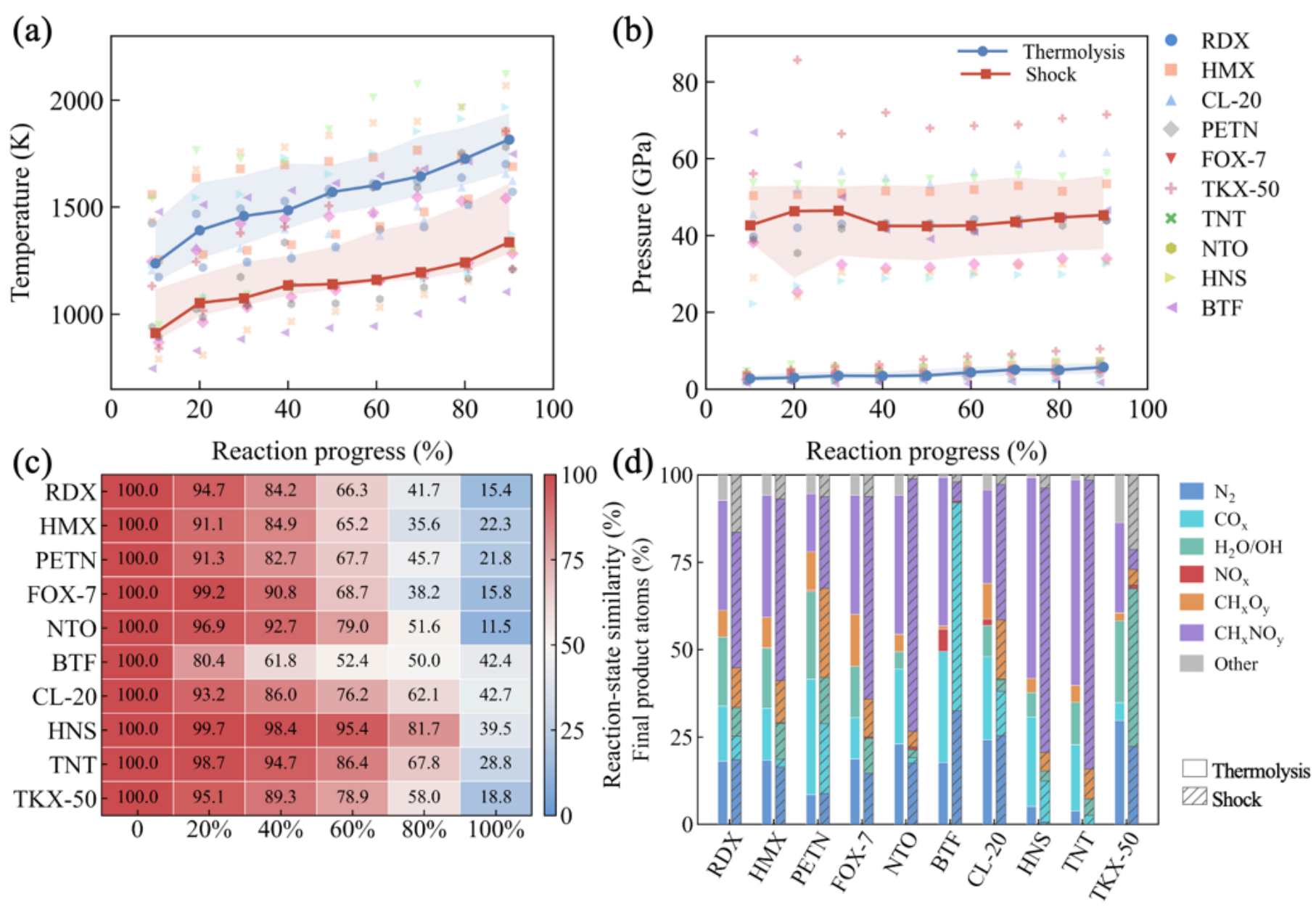

**Fig. 6.** Thermodynamic conditions, reaction-state similarity, and final product composition under thermolysis and shock loading. (a) Temperature and (b) pressure as functions of normalized reaction progress for the 10 energetic materials. Symbols represent individual materials, solid lines indicate the median values across the 10 materials, and shaded areas show the 25$^{th}$-75$^{th}$ percentile ranges. (c) Reaction-state similarity between thermolysis and shock loading for 10 materials. (d) Atomic fractions of the final product families under thermolysis and shock loading, represented by solid and hatched bars, respectively. Products are grouped into $N_2$, $CO_x$, $H_2O/OH$, $NO_x$, $CH_xO_y$, $CH_xNO_y$, and other species.

The degree of thermolysis-shock reaction-state divergence varies among the 10 energetic materials (Fig. 6(c)). BTF exhibits relatively low similarity at early stages, decreasing from 80.4% at 20% reaction progress to 61.8% at 40%. In contrast, HNS retains similarities of 99.7%, 98.4%, and 95.4% at 20%, 40%, and 60% reaction progress, respectively, while TNT maintains 86.4% at 60%. At 100% normalized reaction progress, the similarities range from 11.5% to 42.7% across all 10 materials. NTO, RDX, FOX-7, and TKX-50 exhibit values of 11.5%, 15.4%, 15.8%, and 18.8%, respectively, whereas CL-20, BTF, and HNS retain comparatively higher similarities of 42.7%, 42.4%, and 39.5%. Overall, the evolution and extent of reaction-state divergence are material-dependent, although substantial differences between thermolysis and shock loading persist across all 10 materials.

The final product distributions were also compared on an atomic basis (Fig. 6(d)). Despite the pronounced decline in reaction-state similarity (Fig. 6(c)), $N_2$, $CO_x$, and $H_2O$/OH remain the dominant product families under both loading modes. The combination of similar major product families and divergent intermediate chemical states shows that endpoint composition alone is insufficient to describe the loading-dependent reaction history.

**Expansion of accessible local structural space**

To examine whether the observed chemical-state divergence is accompanied by differences in local atomic environments, we characterized thermolysis and shock trajectories using smooth overlap of atomic positions (SOAP) descriptors. Local environments from thermolysis of 20 materials and shock loading of 10 materials were projected into a unified principal component analysis (PCA) space. SOAP captures the elemental identities and spatial arrangements of neighboring atoms, providing a unified representation of local structural changes associated with bond breaking, atomic rearrangement, and product formation (*56*). The SOAP parameters, joint PCA construction, and structural-distance analysis are detailed in Sections S1.3 and S1.4.

The thermolysis trajectories of the 20 energetic materials occupy partially overlapping regions in the common three-dimensional SOAP-PCA space, with material-specific extensions reflecting differences in their molecular frameworks and initial crystal structures (Fig. 7(a)). The material-resolved PC1-PC2 projections further reveal overlapping distributions with distinct extensions rather than isolated structural regions (Fig. S11). These observations are consistent with the previous EMFF-2025 study, which identified partially overlapping local chemical environments across structurally distinct C/H/N/O energetic materials during high-temperature decomposition (*8*).

For the 10 energetic materials investigated under both loading modes, the shock trajectories extend beyond the corresponding thermolysis distributions in the common SOAP-PCA space (Fig. S12(b)). These additional local environments are distributed along multiple principal-component directions, as illustrated by the shock projection in Fig. 7(b). Together with the material-specific nearest-neighbor analysis, these results show that shock loading accesses local structural environments beyond the corresponding thermolysis reference domains. This expansion of accessible local structural space accompanies the pronounced chemical-state divergence observed at matched reaction progress. The common thermolysis-shock projection and the pressure- and temperature-resolved shock distributions are presented in Fig. S12.

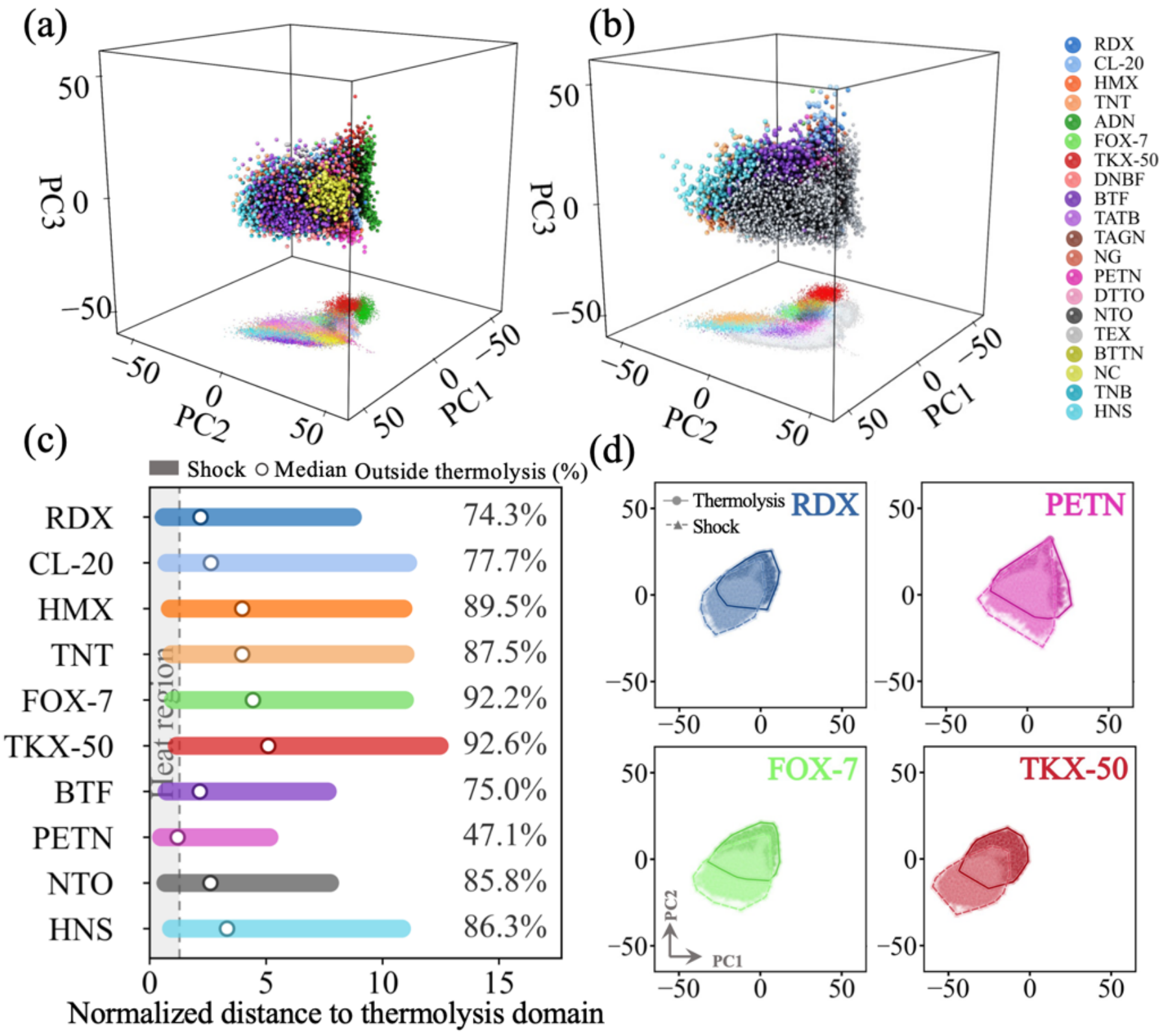

**Fig. 7.** Local structural coverage and shock-induced expansion in a unified SOAP-PCA space. (a) Thermolysis local-environment distribution for 20 energetic materials in the common three-dimensional SOAP-PCA space spanned by PC1, PC2, and PC3. (b) Shock local environments for 10 materials projected into the same space. (c) Normalized distances of shock samples from the material-specific thermolysis reference domains and the corresponding outside-domain fractions. Open circles indicate median distances. (d) Thermolysis and shock structural-domain envelopes of RDX, PETN, FOX-7, and TKX-50 in the PC1-PC2 plane. Quantitative structural-domain distances were evaluated in the full retained PCA representation. The two- and three-dimensional projections are provided for visualization.

The loading-dependent differences in local structural space were quantified using material-specific thermolysis reference domains. For each material, the 95th percentile ($d_{95}$) of the nearest-neighbor distance distribution among thermolysis samples was used as a material-specific reference threshold. The nearest-neighbor distance from each shock sample to the corresponding thermolysis samples was then normalized by d95 to obtain dnorm (Fig. 7(c), Table S6, and Section S1.4). Shock environments with dnorm greater than 1 were classified as lying outside the corresponding thermolysis reference domain. Across the 10 energetic materials, 47.1-92.6% of shock local environments lie outside their corresponding thermolysis reference domains, with a mean fraction of 80.8%. The highest fractions are observed for TKX-50 (92.6%), FOX-7 (92.2%), and HMX (89.5%), whereas PETN exhibits the lowest fraction (47.1%). These results show that shock loading accesses material-dependent local structural environments that are not well represented by the corresponding thermolysis trajectories. Here, the thermolysis reference domain is defined by the sampled trajectories of each material, and environments outside this domain do not necessarily represent extrapolation beyond the EMFF-MACE training domain (Section S1.4).

Material-resolved structural distributions further reveal shock-induced expansion within individual energetic materials (Fig. 7(d)). For RDX, PETN, FOX-7, and TKX-50, the thermolysis and shock distributions partially overlap in the PC1-PC2 plane, while the shock environments extend beyond the corresponding thermolysis envelopes along different directions and to varying extents. Similar material-dependent extensions are observed across all 10 systems (Fig. S13). Together with the nearest-neighbor distance analysis (Fig. 7(c)), these material-resolved comparisons confirm that the additional structural environments accessed under shock loading occur within individual materials, rather than arising solely from pooling data across different systems.

**Intermolecular reorganization and reaction-pathway divergence**

The distinct local structural environments accessed under thermolysis and shock loading raise a further issue concerning the organization and connectivity of reactive intermediates. Previous atomistic studies have highlighted the roles of molecular strain and local mechanical environments in shock-induced chemical activation (*13, 14*). Here, we extend the comparison beyond initial activation to examine intermolecular proximity, connectivity, and subsequent reaction-network evolution. RDX, PETN,

FOX-7, and TKX-50 were selected to examine loading-dependent intermolecular organization and structural evolution. Inter-component radial distributions are illustrated for PETN in Fig. 8(a), with corresponding results for RDX, FOX-7, and TKX-50 provided in Fig. S15. Their mean structural trajectories at matched normalized reaction progress are compared in Fig. 8(b). Among the 10 energetic materials, PETN and TKX-50 exhibit the lowest (47.1%) and highest (92.6%) fractions of shock local environments outside their respective thermolysis reference domains (Fig. 7(c)), and were therefore selected for detailed mechanical and reaction-network analyses (Fig. S14 and Fig. 9).

Parent-unit-resolved virial analysis reveals distinct local mechanical responses of PETN and TKX-50 under thermolysis and shock loading (Fig. S14). Both materials exhibit higher hydrostatic and von Mises virial components under shock loading, reflecting enhanced compressive and deviatoric contributions. The hydrostatic component is particularly pronounced in TKX-50, indicating material-dependent differences in the local mechanical response. These results provide a mechanical context for the loading-dependent intermolecular reorganization examined below.

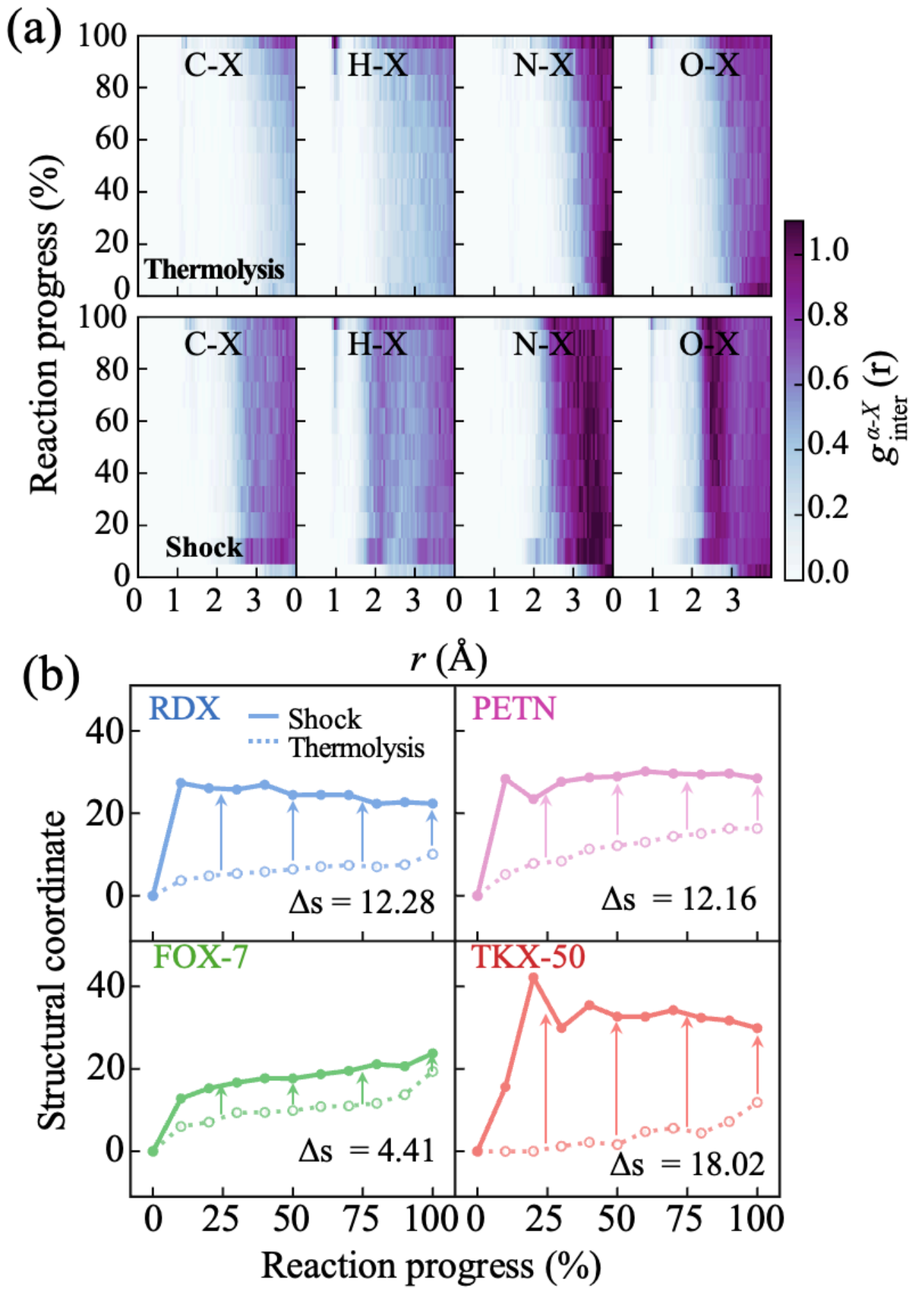

**Fig. 8.** Inter-component atomic distributions and structural-coordinate separation under thermolysis

and shock loading. (a) Inter-component radial distribution functions of PETN under thermolysis (upper panels) and shock loading (lower panels). C–X, H–X, N–X, and O–X denote inter-component atomic pairs, where X represents any C/H/N/O atom belonging to another initial molecular component. The color scale represents the corresponding inter-component radial distribution function, $g(r)$. (b) SOAP-PCA comparison of thermolysis and shock structural evolution at matched normalized reaction progress. Δs denotes the separation between their mean projected structural coordinates along a material-specific common axis.

Inter-component radial distributions reveal loading-dependent changes in intermolecular organization (Fig. 8(a)). For PETN, shock loading enhances short-range correlations for all four atomic-pair types (C–X, H–X, N–X, and O–X), with particularly pronounced increases in N–X and O–X. These changes persist across much of the normalized reaction-progress range and indicate increased proximity between atoms from initially separate molecular components. Similar loading-dependent changes are observed in RDX, FOX-7, and TKX-50 (Fig. S15). Complementary close-contact and large-cluster analyses further characterize the differences in intermolecular proximity and connectivity (Fig. S16). SOAP-PCA analysis also reveals loading-dependent differences in mean structural evolution at matched normalized reaction progress (Fig. 8(b)). At 100% normalized reaction progress, the structural-coordinate separations (Δs) are 12.28, 12.16, 4.41, and 18.02 for RDX, PETN, FOX-7, and TKX-50, respectively (Table S7). Together, these results reveal changes in intermolecular proximity, connectivity, and structural evolution that accompany the loading-dependent differences in chemical states.

The representative PETN reaction network shows that thermolysis and shock loading share early nitrate-ester cleavage processes and $NO_2$-containing intermediates (Fig. 9(a)). Their subsequent transformations, however, follow different branches. Under thermolysis, successive denitration produces progressively smaller PETN-derived fragments. The shock network additionally contains net transformations involving initially separate molecular components, including the formation of larger $C_{10}$-containing structures and associated rearrangement branches. These intermolecular association branches are observed in the selected shock network but not in the corresponding thermolysis branches. Thus, the two loading modes share initial nitrate-ester decomposition chemistry but differ in the subsequent connectivity of PETN-derived intermediates. The additional $C_{10}$-containing structures provide a specific chemical example of the loading-dependent pathway divergence associated with intermolecular reorganization.

TKX-50 exhibits a related pathway distinction despite its different initial ionic structure (Fig. 9(b)). Both loading modes involve hydrogen-transfer reactions and the formation of $H_2O$-containing products. Under thermolysis, the representative pathway proceeds through successive hydrogen transfer, $H_2O$ elimination, and fragmentation of the original ionic components. Under shock loading, additional association and rearrangement reactions involve initially separate nitrogen-rich components, producing larger connected intermediates and alternative fragmentation branches. These results extend the observed pathway distinction to an ionic energetic material: early decomposition chemistry partly overlaps between the two loading modes, whereas shock loading is associated with additional connectivity and rearrangement involving initially separate components.

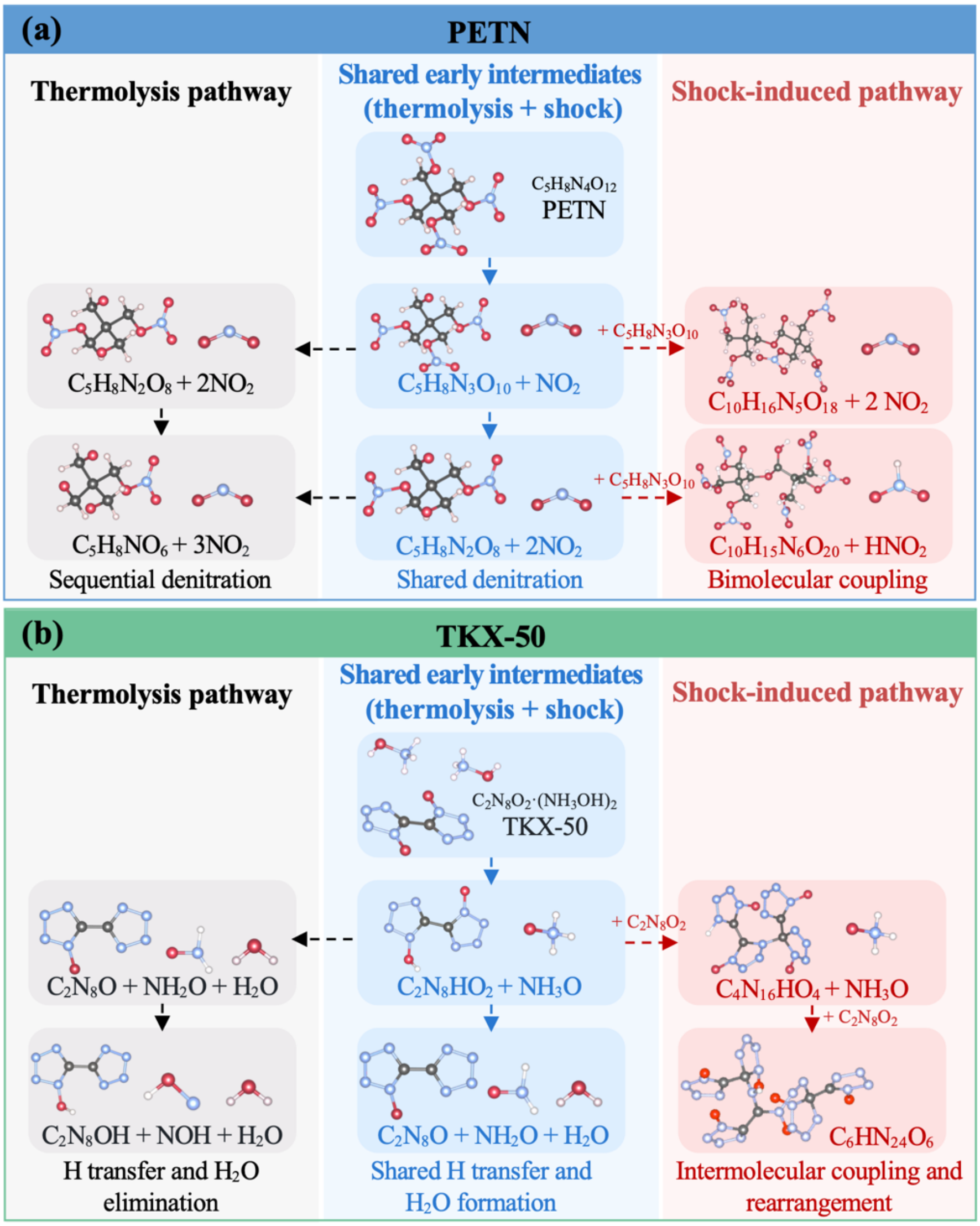

**Fig. 9.** Representative reaction networks of (a) PETN and (b) TKX-50 under thermolysis and shock loading. Blue panels indicate shared early intermediates and transformations, gray panels represent thermolysis branches, and red panels highlight additional shock-associated branches. Gray, white, blue, and red spheres represent C, H, N, and O atoms, respectively. The networks were constructed from atom-mapped, element-conserving net transformations identified over consecutive 1 ps

analysis intervals.

The principal distinction between thermolysis and shock loading emerges in the reactions following initial molecular decomposition. For PETN and TKX-50 (Fig. 9), both loading modes exhibit partially shared early decomposition pathways, whereas shock loading introduces additional intermolecular association and rearrangement branches, accompanied by the formation of larger connected intermediates. Similar loading-dependent pathway divergence is observed for RDX and FOX-7 (Fig. S18), where shock loading exhibits additional intermolecular connectivity and distinct subsequent reaction branches. Related shock-induced and mechanochemical reaction mechanisms have also been discussed in previous studies (*25-27*). Together, these results demonstrate that shock-induced chemical evolution involves changes in molecular connectivity and reaction-pathway selection, rather than merely accelerated thermolysis.

# Discussion

The common EMFF-MACE potential energy surface enables a consistent comparison of thermolysis and shock-induced chemistry across diverse C/H/N/O energetic materials. Validation against first-principles energies and forces, crystal properties, and detonation-state benchmarks supports the use of EMFF-MACE over the configurations and conditions evaluated here.

A central finding is that neither normalized reaction progress nor major stable product composition uniquely characterizes chemical evolution across different loading histories. Thermolysis and shock loading exhibit increasingly distinct intermediate chemical states and access different local structural environments, despite partially shared decomposition chemistry. This distinction indicates that the history of thermal and mechanical loading is an important dimension of reaction-pathway evolution.

These comparisons have several limitations. Reaction trajectories were aligned at matched normalized reaction progress, not at identical temperature, pressure, or absolute parent-reactant conversion. The observed differences therefore reflect the combined effects of the loading histories rather than independently isolated contributions from compression, heating, and deformation rate. In addition, the structural reference domains are defined by the sampled thermolysis trajectories, and the representative reaction networks capture net changes in molecular connectivity over finite analysis intervals rather than uniquely resolved elementary steps. These considerations delimit the interpretation of the observed structural and pathway differences.

The observed reaction-pathway differences are accompanied by pronounced intermolecular reorganization under shock loading. Although early fragmentation and hydrogen-transfer processes partly overlap, the PETN and TKX-50 reaction networks exhibit additional intermolecular association and rearrangement branches under shock loading, together with enhanced inter-component proximity and distinct local mechanical environments. These findings identify intermolecular reorganization and changes in molecular connectivity as structural features associated with loading-dependent pathway divergence. Shock-induced chemistry therefore differs from thermolysis not simply in reaction rate, but also in the structural states and subsequent reaction pathways accessed during decomposition.

## Materials and Methods

### Dataset construction and EMFF-MACE training

The EMFF-MACE training data were constructed from the previously developed EMFF-2025 dataset (*8*) and covered 20 C/H/N/O energetic materials (Fig. 2(a)): RDX, HMX, CL-20, TNT, ADN, FOX-7, TKX-50, DNBF, BTF, TATB, TAGN, NG, PETN, DTTO, NTO, TEX, BTTN, NC, TNB, and HNS (Table S1). The dataset spans 300-4000 K and includes near-equilibrium, high-temperature, and reactive configurations. To enhance the representation of high-density and strongly deformed atomic environments relevant to shock loading, the original EMFF-2025 dataset was augmented with 27,000 additional configurations for RDX, HMX, and CL-20, extending the sampled linear cell-scaling range from 0.92-1.08 to 0.80-1.20. Together with the original 33,900 configurations, the resulting dataset comprises 60,900 configurations, which were randomly divided into training and test subsets at a ratio of 9:1, yielding 54,810 and 6,090 configurations, respectively (Table S4). All configurations were evaluated using DFT calculations with CP2K (*57*), employing the Perdew-Burke-Ernzerhof (PBE) generalized-gradient approximation (*58*) together with the D3 dispersion correction (DFT-D3) (*59*). Total energies and atomic forces were used as reference labels for training and evaluation. Additional details of the original dataset construction and first-principles calculations are provided in our previous

EMFF-2025 study (*8*).

EMFF-MACE was constructed using the higher-order equivariant MACE architecture (*28*). Training and LAMMPS coupling were implemented through DeePMD-GNN, a graph neural network (GNN) plugin within the DeePMD-kit v3 framework (*60-62*). MACE represents local atomic environments using an E(3)-equivariant graph representation and captures many-body interactions through higher-order feature correlations. The EMFF-MACE model was constructed for four elements (C, H, N, and O), with a local-environment cutoff radius of 6.0 Å. The hidden features were configured with 64 scalar channels of even parity (angular momentum $\ell = 0$), denoted as $64 \times 0e$ in the e3nn notation. Model training employed a joint loss function incorporating total energies and atomic forces. During training, the energy loss weight was gradually increased from 0.02 to 1, while the force loss weight was decreased from 1000 to 1. The learning rate was exponentially decayed from $1 \times 10^{-3}$ to $1 \times 10^{-8}$, with updates applied every 500 optimization steps. Training was performed for $10^6$ steps. A single trained potential was subsequently applied to crystal-structure, thermolysis, and shock simulations of C/H/N/O energetic materials.

**Validation and simulation protocols**

Thermolysis and MSST shock simulations were performed using the same EMFF-MACE potential energy surface (Fig. 2(b)), providing a common computational basis for comparing the two loading modes. Chemical species were identified using consistent criteria, and trajectories were aligned by normalized reaction progress, defined relative to the total parent-reactant depletion reached in each trajectory (Section S1.1). Species distributions, SOAP-based local structural descriptors, inter-component atomic distributions, and clustering statistics were then analyzed to characterize loading-dependent differences in chemical states and intermolecular organization.

The accuracy of EMFF-MACE was evaluated against DFT reference energies and atomic forces (Fig. 2(c)). Additional validation included temperature-dependent and deformed configurations, equilibrium crystal structures, equations of state, and representative internal-coordinate potential-energy scans. MAE, RMSE, and $R^2$ were calculated for the test-set energies and forces, with prediction errors further examined across all 20 materials. For RDX, HMX, and CL-20, the evaluation covered temperatures of 300, 1000, 2000, 3000, and 4000 K and isotropic cell-scaling factors of 0.80-1.20. Crystal structures and equations of state were benchmarked against DFT calculations and available experimental data. First-principles calculations were performed using CP2K (*57*) with the PBE exchange-correlation functional (*58*) and DFT-D3 dispersion correction (*59*), employing the DZVP-MOLOPT basis set, a plane-wave cutoff of 400 Ry, a relative cutoff of 60 Ry, and a self-consistent field convergence threshold of $1 \times 10^{-6}$. Representative bond-length, bond-angle, and dihedral-angle scans were additionally performed for RDX and PETN to evaluate the local potential-energy response.

All molecular dynamics simulations were performed using LAMMPS (*61*) coupled with the same EMFF-MACE potential. Thermolysis simulations employed three-dimensional periodic boundary conditions, with each system containing at least 32 energetic molecules or equivalent structural units. After initial equilibration in the canonical (NVT) ensemble at 300 K, the systems were linearly heated from 300 to 3000 K over 120 ps with a time step of 0.2 fs, corresponding to a heating rate of 22.5 $K \cdot ps^{-1}$. The same species-identification and product-analysis procedures were applied to all 20 materials. The crystal forms, initial densities, and loading directions of the 10 shock-loaded energetic materials are summarized in Table S5.

Shock simulations were conducted using MSST (*34*), which couples atomic motion with dynamic cell deformation subject to the Rankine-Hugoniot conservation relations. Shock propagation was defined along the *z* axis of the simulation cell. For PETN, ε-CL-20, and *trans*-HNS, the crystallographic [100], [001], and [010] directions, respectively, were aligned with the simulation *z* axis before shock loading. For the remaining materials, shock loading was applied along the z axis of their relaxed simulation cells without additional crystallographic reorientation. Each shock trajectory was propagated for 120 ps, and the time evolution of relative volume ($V/V_0$), temperature ($T$), pressure ($P$), and density ($\rho$) was recorded. For RDX, shock velocities from 6.0 to 10.0 $km \cdot s^{-1}$ were sampled, with additional sampling near the reaction-transition regime. The same MSST protocol was applied to HMX, PETN, FOX-7, NTO, BTF, CL-20, HNS, TNT, and TKX-50. The CJ state was determined from dynamically stabilized post-reaction states and the tangency between the Rayleigh line and the reacted Hugoniot (*5, 34, 35*), yielding $D_{CJ}$, $P_{CJ}$, $T_{CJ}$, and $\rho_{CJ}$.

**Statistical analysis**

Prediction errors were evaluated on the randomly selected test subset of 6,090 DFT-labeled configurations using MAE, RMSE, and $R^2$. Reaction-state similarities were calculated from overlapping normalized species-abundance distributions at matched normalized reaction progress (Section S1.2). For the cross-material temperature and pressure summaries in Figs. 6(a)-(b), solid curves show the median across ten materials and shaded bands show the $25^{th}$-$75^{th}$ percentile interval.

## Supplementary Materials

This PDF file includes:

Supplementary Text

Figs. S1 to S18

Tables S1 to S7

## Acknowledgments

**Funding:** This work was supported by the National Natural Science Foundation of China (grant nos. 52676271, 52106130, and 22675032) and the State Key Laboratory of Explosion Science and Safety Protection (grant no. QNKT23-15).

**Author contributions:** Conceptualization: M.W. and D.C. Methodology: M.W. and Q.C. Model development and simulations: M.W. and J.H. Data analysis and visualization: M.W., J.H., Z.Z., and H.W. Validation and interpretation: M.W. and X.G. Writing—original draft: M.W. Writing—review and editing: Q.C. and D.C. Supervision, project administration, resources, and funding acquisition: Q.C. and D.C. All authors discussed the results and approved the manuscript.

**Competing interests:** The authors declare that they have no competing interests.

**Data, code, and materials availability:** The trained EMFF-MACE potential, simulation inputs, analysis scripts, and data supporting the findings of this study are available at [https://github.com/MingjieWen/General-NNP-model-for-C-H-N-O-Energetic-Materials/tree/main]. No new physical materials were generated in this computational study.